\documentstyle[prd,preprint,tighten,aps,amssymb,newlfont,epsfig]{revtex}
\begin{document}
\draft
\preprint{
\begin{tabular}{r}
   UWThPh-1999-26
\\ DFTT 21/99
\\
\end{tabular}
}

\title{Neutrino oscillation constraints
on neutrinoless double-beta decay}

\author{S.M. Bilenky}
\address{Joint Institute for Nuclear Research, Dubna, Russia, and\\
Institute for Theoretical Physics, University of Vienna,\\
Boltzmanngasse 5, A--1090 Vienna, Austria}
\author{C. Giunti}
\address{INFN, Sez. di Torino, and Dip. di Fisica Teorica,
Univ. di Torino,\\
Via P. Giuria 1, I--10125 Torino, Italy}

\maketitle

\begin{abstract}
It is shown that,
in the framework of the scheme with three-neutrino mixing
and a mass hierarchy,
the results of neutrino oscillation experiments
imply an upper bound of about $10^{-2}$ eV
for the effective Majorana mass in neutrinoless double-$\beta$ decay.
The schemes with four massive neutrinos are also briefly discussed.
\end{abstract}

\pacs{Talk presented by C. Giunti at the
17$^{\mathrm{th}}$ \textit{International Workshop on Weak Interactions and Neutrinos}
(WIN99),
Cape Town, South Africa, 24--30 January 1999.}

The recent results of the high-precision and high-statistics
Super-Kamiokande experiment\cite{SK-atm,SK-sun}
have confirmed the indications
in favor of neutrino
oscillations\cite{BGG-review} obtained in
atmospheric\cite{atm-exp} and solar\cite{sun-exp} neutrino experiments.
Here we will discuss
the implications of the results of atmospheric and solar neutrino
oscillation experiments for neutrinoless double-$\beta$ decay
($(\beta\beta)_{0\nu}$)
in the framework of the scheme with three neutrinos and a mass hierarchy,
that can accommodate atmospheric and solar neutrino oscillations,
and in the framework of the schemes with four massive neutrinos
that can accommodate also the
$\bar\nu_\mu\to\bar\nu_e$
and
$\nu_\mu\to\nu_e$
oscillations observed in the LSND
experiment.\cite{LSND}

The results of atmospheric neutrino experiments can be explained
by $\nu_\mu\to\nu_\tau$ oscillations
due to the mass-squared difference\cite{SK-atm}
\begin{equation}
\Delta{m}^2_{\mathrm{atm}}
\sim
( 1 - 8 ) \times 10^{-3} \, \mathrm{eV}^2
\,.
\label{dm2atm}
\end{equation}
The results of solar neutrino experiments can be explained by
$\nu_e\to\nu_\mu,\nu_\tau$ transitions
due to the mass-squared difference
\begin{equation}
\Delta{m}^2_{\mathrm{sun}}
\sim
( 0.5 - 10 ) \times 10^{-10} \, \mathrm{eV}^2
\qquad
\mbox{(VO)}
\label{dm2sun-VO}
\end{equation}
in the case of vacuum oscillations,
or
\begin{equation}
\Delta{m}^2_{\mathrm{sun}}
\sim
( 0.4 - 1 ) \times 10^{-5} \, \mathrm{eV}^2
\qquad
\mbox{(SMA-MSW)}
\label{dm2sun-SMA}
\end{equation}
in the case of small mixing angle MSW transitions,
or
\begin{equation}
\Delta{m}^2_{\mathrm{sun}}
\sim
( 0.6 - 20 ) \times 10^{-5} \, \mathrm{eV}^2
\qquad
\mbox{(LMA-MSW)}
\label{dm2sun-LMA}
\end{equation}
in the case of large mixing angle MSW transitions.\cite{sun-analysis}
Hence,
atmospheric and solar neutrino data indicate a hierarchy of $\Delta{m}^2$'s:
$ \Delta{m}^2_{\mathrm{sun}}
\ll
\Delta{m}^2_{\mathrm{atm}}
$.
A natural scheme that can accommodate this hierarchy
is the one with three neutrinos and a mass hierarchy
$ m_1 \ll m_2 \ll m_3 $,
that is predicted by the see-saw mechanism.
In this case we have
\begin{equation}
\Delta{m}^2_{\mathrm{sun}}
=
\Delta{m}^2_{21}
\simeq
m_2^2
\,,
\qquad
\Delta{m}^2_{\mathrm{atm}}
=
\Delta{m}^2_{31}
\simeq
m_3^2
\,.
\label{dm2-hierarchy}
\end{equation}

In the spirit of the see-saw mechanism,
we presume that massive neutrinos are Majorana particles
and 
$(\beta\beta)_{0\nu}$-decay
is allowed.
The matrix element
of $(\beta\beta)_{0\nu}$ decay is proportional to the effective Majorana
neutrino mass
\begin{equation}
|\langle{m}\rangle|
=
\left|
\sum_{k}
U_{ek}^2
\,
m_{k}
\right|
\,,
\label{effective}
\end{equation}
where $U$ is the mixing matrix
that connects the flavor neutrino fields
$\nu_{\alpha L}$
($\alpha=e,\mu,\tau$)
to the fields $\nu_{kL}$ of neutrinos with masses $m_k$
through the relation
$ \nu_{\alpha L} = \sum_{k} U_{\alpha k} \nu_{kL} $.
The present experimental upper limit for $|\langle{m}\rangle|$
is 0.2 eV at 90\% CL.\cite{Baudis-99}
The next generation of $(\beta\beta)_{0\nu}$ decay experiments
is expected to be sensitive to values of
$|\langle{m}\rangle|$
in the range $10^{-2} - 10^{-1}$ eV.

Since the results of neutrino oscillation experiments
allow to constraint only the moduli of the elements
$U_{ek}$ of the neutrino mixing matrix,
let us consider the upper bound
\begin{equation}
|\langle{m}\rangle|
\leq
\sum_{k}
|U_{ek}|^2
\,
m_{k}
\equiv
|\langle{m}\rangle|_{\mathrm{UB}}
\,.
\label{bound}
\end{equation}

In the framework of the scheme with three neutrinos and
a mass hierarchy,
the contribution of $m_1$ to $|\langle{m}\rangle|_{\mathrm{UB}}$
is negligible
and
the contributions of $m_2$ and $m_3$
are given, respectively, by
\begin{eqnarray}
&&
|\langle{m}\rangle|_{\mathrm{UB}2}
\equiv
|U_{e2}|^2
\,
m_{2}
\simeq
|U_{e2}|^2
\,
\sqrt{ \Delta{m}^2_{\mathrm{sun}} }
\label{m2}
\\
&&
|\langle{m}\rangle|_{\mathrm{UB}3}
\equiv
|U_{e3}|^2
\,
m_{3}
\simeq
|U_{e3}|^2
\,
\sqrt{ \Delta{m}^2_{\mathrm{atm}} }
\,.
\label{m3}
\end{eqnarray}

The parameter $|U_{e2}|^2$
is large
in the case of solar vacuum oscillations,
is smaller than $1/2$
in the case of large mixing angle MSW transitions
and
is very small ($|U_{e2}|^2 \lesssim 2 \times 10^{-3}$)
in the case of small mixing angle MSW transitions.
Taking into account the respective ranges of $\Delta{m}^2_{\mathrm{sun}}$
in Eqs.(\ref{dm2sun-VO})--(\ref{dm2sun-LMA}),
we have
\begin{equation}
|\langle{m}\rangle|_{\mathrm{UB}2}
\lesssim
\left\{
\begin{array}{lcl}
3 \times 10^{-5} \, \mathrm{eV}
& \quad &
\mbox{(VO)}
\,,
\\
6 \times 10^{-6} \, \mathrm{eV}
& \quad &
\mbox{(SMA-MSW)}
\,,
\\
7 \times 10^{-3} \, \mathrm{eV}
& \quad &
\mbox{(LMA-MSW)}
\,.
\end{array}
\right.
\label{m2-sun}
\end{equation}
Hence,
the contribution of $m_2$ to the upper bound (\ref{bound})
is small
and one expects that
the dominant contribution is given by
$ m_3
\simeq
\sqrt{ \Delta{m}^2_{\mathrm{atm}} }
\sim
( 3 - 9 ) \times 10^{-2} \, \mathrm{eV}
$.
However,
as we will show in the following,
the value
of $|U_{e3}|^2$ is constrained by the results of the atmospheric
Super-Kamiokande experiment
and by the negative results of the long-baseline reactor
$\bar\nu_e$ disappearance experiment CHOOZ.\cite{CHOOZ-98}

The two-neutrino exclusion plot obtained in the CHOOZ experiments
imply that\cite{BGKM98-bb}
$ |U_{e3}|^2 \leq a_e^{\mathrm{CHOOZ}} $
or
$ |U_{e3}|^2 \geq 1-a_e^{\mathrm{CHOOZ}} $,
with
$ a_e^{\mathrm{CHOOZ}}
=
\frac{1}{2} \left( 1 - \sqrt{ 1 - \sin^2 2\vartheta_{\mathrm{CHOOZ}} } \right)
$.
Here $\sin^2 2\vartheta_{\mathrm{CHOOZ}}$
is the upper value of the two-neutrino mixing parameter
$\sin^2 2\vartheta$
obtained from the CHOOZ exclusion curve
as a function of
$\Delta{m}^2 = \Delta{m}^2_{31} = \Delta{m}^2_{\mathrm{atm}}$,
where $\Delta{m}^2$ is the two-neutrino mass-squared difference.
Since the quantity $a_e^{\mathrm{CHOOZ}}$ is small
for
$\Delta m^2_{\mathrm{atm}} \gtrsim 10^{-3} \, \mathrm{eV}^2$,
the results of the CHOOZ experiment imply that
$|U_{e3}|^2$ is either small or close to one.
However,
since the survival probability of solar $\nu_e$'s is bigger than
$|U_{e3}|^4$,
only the range
$|U_{e3}|^2 \leq a_e^{\mathrm{CHOOZ}}$
is allowed by the results of solar neutrino experiments.
Therefore,
the contribution of $m_3$ to $|\langle{m}\rangle|_{\mathrm{UB}}$
is bounded by
\begin{equation}
|\langle{m}\rangle|_{\mathrm{UB}3}
\lesssim
a_e^{\mathrm{CHOOZ}} \, \sqrt{ \Delta{m}^2_{\mathrm{atm}} }
\,.
\label{m3-bound}
\end{equation}
Notice that this limit depends on
$\Delta{m}^2_{\mathrm{atm}}$
both explicitly and implicitly through $a_e^{\mathrm{CHOOZ}}$.

\begin{figure}[h]
\begin{center}
\epsfig{file=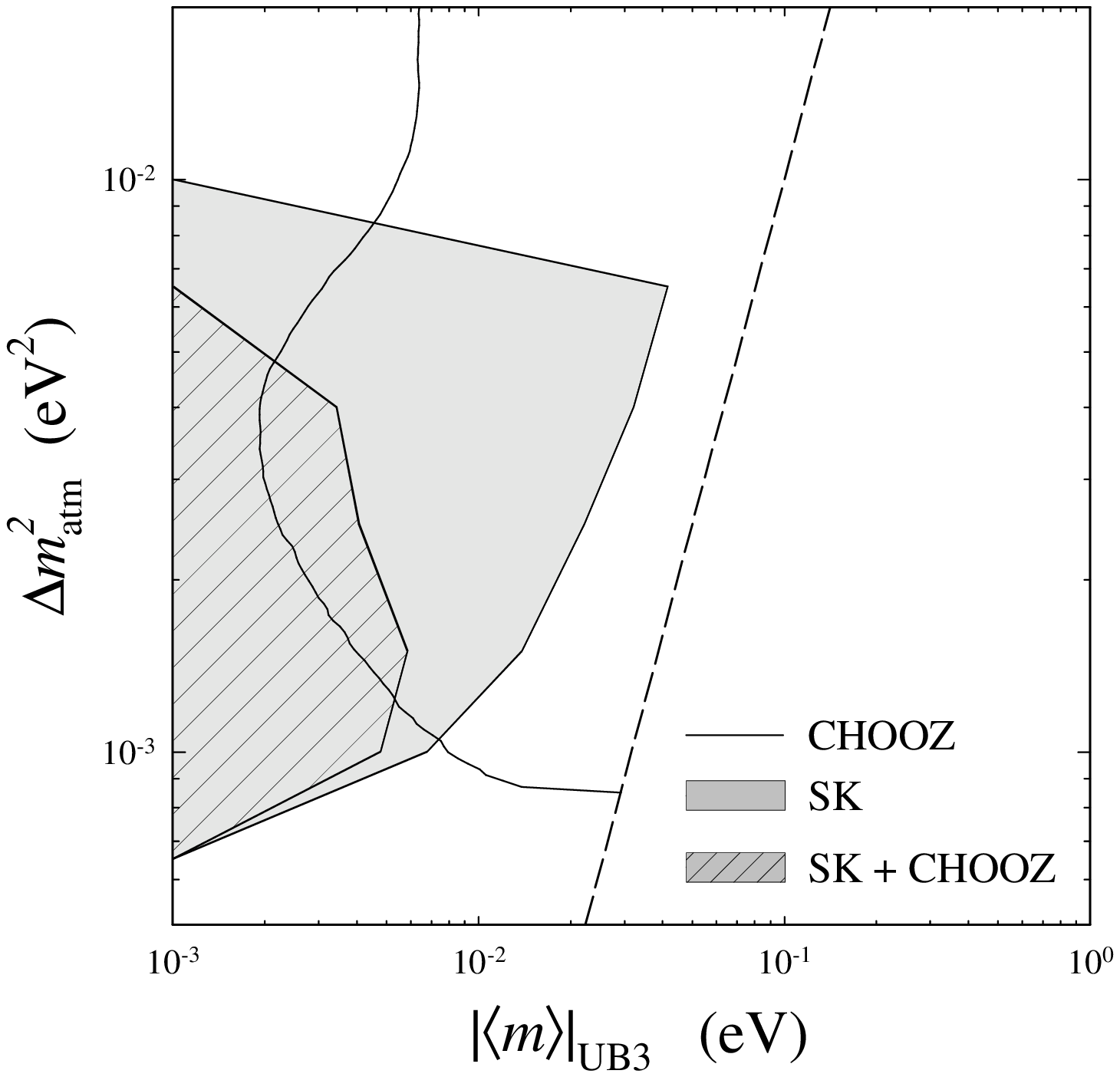,width=0.50\linewidth}
\caption{ \label{bb3} }
\end{center}
\end{figure}

The bound
in the $|\langle{m}\rangle|_{\mathrm{UB}3}$--$\Delta{m}^2_{\mathrm{atm}}$
plane obtained from the inequality (\ref{m3-bound})
using the CHOOZ exclusion curve
is shown in Fig.~\ref{bb3}
by the solid line
(the region on the right of this curve is excluded).
The dashed straight line in Fig.~\ref{bb3}
represents the unitarity bound
$ |\langle{m}\rangle|_{\mathrm{UB}3}
\leq \sqrt{ \Delta{m}^2_{\mathrm{atm}} }
$.

The shadowed and hatched regions
in Fig.~\ref{bb3}
are allowed\cite{FLMS98} by the analysis of the Super-Kamiokande data and
the combined analysis of the Super-Kamiokande and CHOOZ data,
respectively.
One can see that the value of $|\langle{m}\rangle|_{\mathrm{UB}3}$
is tightly constrained:
\begin{equation}
|\langle{m}\rangle|_{\mathrm{UB}3}
\lesssim
6 \times 10^{-3} \, \mathrm{eV}
\,.
\label{m3-limit}
\end{equation}
Therefore,
taking into account the inequalities (\ref{bound}),
(\ref{m2-sun})
and
(\ref{m3-limit}),
we conclude that in the scheme with three neutrinos and a mass hierarchy
the effective Majorana mass $|\langle{m}\rangle|$
in $(\beta\beta)_{0\nu}$-decay
is bounded by
\begin{equation}
|\langle{m}\rangle|
\lesssim
10^{-2} \, \mathrm{eV}
\,.
\label{limit}
\end{equation}

Let us consider now the two schemes with four-neutrino mixing
that can accommodate the results of
solar and atmospheric neutrino experiments
and the results of the accelerator LSND experiment:\cite{BGG-AB}
\begin{equation}
\mbox{(A)}
\qquad
\underbrace{
\overbrace{m_1 < m_2}^{\mathrm{atm}}
\ll
\overbrace{m_3 < m_4}^{\mathrm{sun}}
}_{\mathrm{LSND}}
\,,
\qquad
\mbox{(B)}
\qquad
\underbrace{
\overbrace{m_1 < m_2}^{\mathrm{sun}}
\ll
\overbrace{m_3 < m_4}^{\mathrm{atm}}
}_{\mathrm{LSND}}
\,.
\label{AB}
\end{equation}
These two spectra are characterized by the presence of two couples
of close masses separated by a gap of about 1 eV
which provides the mass-squared difference
$ \Delta{m}^2_{\mathrm{LSND}} = \Delta{m}^2_{41} $
responsible of the oscillations observed in the LSND experiment.
In the scheme A
$ \Delta{m}^2_{\mathrm{atm}} = \Delta{m}^2_{21} $
and
$ \Delta{m}^2_{\mathrm{sun}} = \Delta{m}^2_{43} $,
whereas in scheme B
$ \Delta{m}^2_{\mathrm{atm}} = \Delta{m}^2_{43} $
and
$ \Delta{m}^2_{\mathrm{sun}} = \Delta{m}^2_{21} $.

\begin{figure}[h]
\begin{center}
\epsfig{file=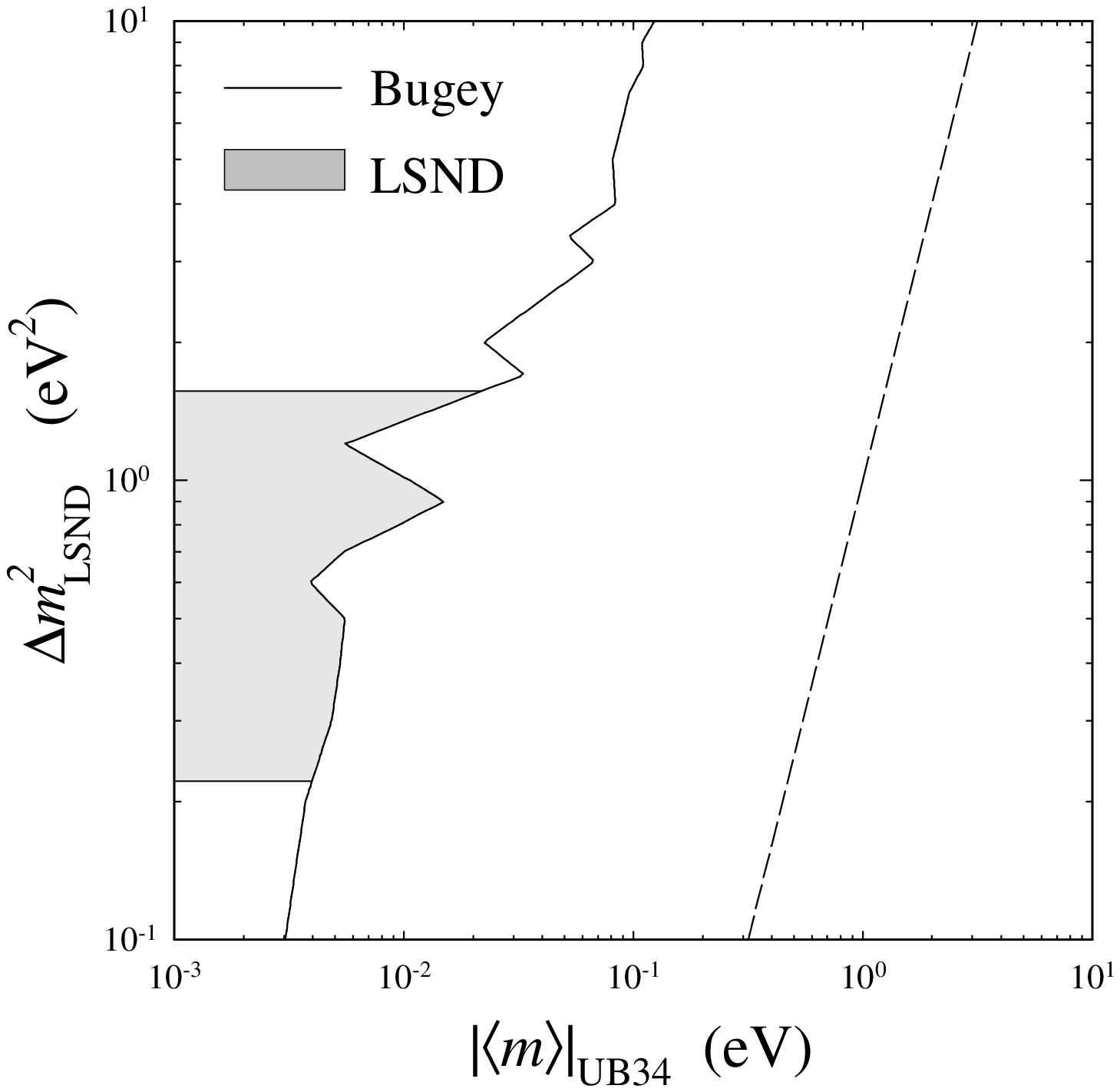,width=0.50\linewidth}
\caption{ \label{bb4} }
\end{center}
\end{figure}

It has been shown\cite{BGG-AB} that the results of
the short-baseline $\bar\nu_e$ disappearance experiment Bugey,\cite{Bugey95}
in which no indication in favor of neutrino oscillations was found,
imply that the mixing
of $\nu_e$ with the two
``heavy'' neutrinos $\nu_3$ and $\nu_4$
is large in scheme A
and
small in scheme B.
Therefore,
if scheme A is realized in nature
the effective Majorana mass in
$(\beta\beta)_{0\nu}$ decay
can be as large as
$ m_3 \simeq m_4 \simeq \sqrt{ \Delta{m}^2_{\mathrm{LSND}} }
\simeq 0.5 - 1.2 \, \mathrm{eV} $.
On the other hand,
in scheme B
$(\beta\beta)_{0\nu}$ decay
is strongly suppressed.
Indeed,
the contribution of
$m_2$ to the upper bound (\ref{bound}) is
limited by Eq.(\ref{m2-sun})
and the contribution of $m_3$ and $m_4$,
$ |\langle{m}\rangle|_{\mathrm{UB}34}
\simeq
( |U_{e3}|^2 + |U_{e4}|^2 ) \sqrt{ \Delta{m}^2_{\mathrm{LSND}} } $,
is limited by the inequality
\begin{equation}
|\langle{m}\rangle|_{\mathrm{UB}34}
\lesssim
a_e^{\mathrm{Bugey}} \, \sqrt{ \Delta{m}^2_{\mathrm{LSND}} }
\,,
\label{B4-bound}
\end{equation}
where
$ a_e^{\mathrm{Bugey}} $
is given by the exclusion curve of the Bugey experiment.
The numerical value of the upper bound (\ref{B4-bound})
is depicted in Fig.~\ref{bb4}
by the solid line.
The dashed straight line in Fig.~\ref{bb4}
represents the unitarity bound
$ |\langle{m}\rangle|_{\mathrm{UB}34}
\leq \sqrt{ \Delta{m}^2_{\mathrm{LSND}} }
$
and the shadowed region indicates the
interval of $\Delta{m}^2_{\mathrm{LSND}}$
allowed at 90\% CL by the results of the LSND experiment:
$ 0.22 \, \mathrm{eV}^2
\leq \Delta{m}^2_{\mathrm{LSND}} \leq
1.56 \, \mathrm{eV}^2 $.
From Fig.~\ref{bb4} one can see that
$ |\langle{m}\rangle|_{\mathrm{UB}34}
\lesssim 2 \times 10^{-2} \, \mathrm{eV}
$.
Therefore, in scheme B we have the upper bound
\begin{equation}
|\langle{m}\rangle|
\lesssim
2 \times 10^{-2} \, \mathrm{eV}
\,.
\label{B4-limit}
\end{equation}

In conclusion,
the results of the analysis of neutrino oscillation data
show that the effective Majorana mass
$|\langle{m}\rangle|$
in neutrinoless double-$\beta$ decay is
smaller than about $10^{-2}$ eV
in the scheme with mixing of three neutrinos
and a mass hierarchy,
is smaller than about $2 \times 10^{-2}$ eV
in the four-neutrino mixing scheme B,
whereas it can be as large as
$ \sqrt{ \Delta{m}^2_{\mathrm{LSND}} }
\simeq 0.5 - 1.2 \, \mathrm{eV} $
in the four-neutrino mixing scheme A.

\end{document}